\documentclass[11pt]{article} 
\usepackage{hyperref} 
\pdfoutput=1 
 
\begin{document} 
 
\title{Flight of a Fly} 
 
\author{H. Aryafar \& H. P. Kavehpour \\ 
\\\vspace{6pt} Department of Mechanical and Aerospace 
Engineering, \\ University of California, Los Angeles \\ Los Angeles, CA 90095, USA} 
 
\maketitle 
 
%% The abstract (in this file, and that submitted as text to arXiv) should include the exact phrase 
%% "fluid dynamics video" or "fluid dynamics videos" 
 
\begin{abstract} 
In this fluid dynamics video, we demonstrated take off and landing of a fly. The deformation of wings is in focus in this video.
\end{abstract} 
 
% main text 
 
\section{Introduction}

We have recorded the flight of a fly during take off and landing using digital high speed photography. It is shown that the dynamics of flexible wings are different for these procedures. During this observation fly flew freely in a big box and it was not tethered.

\end{document}